\documentclass{article}
\usepackage{graphicx,cite,bm}
\newcommand{\PRL}{\textit{Phys. Rev. Lett. }}
\newcommand{\PRB}{\textit{Phys. Rev. B }}

\begin{document}
\title{Structures, Symmetries, Mechanics and Motors of carbon nanotubes}
\author{Z. C. Tu and Z. C. Ou-Yang\\
\textit{Institute of Theoretical Physics,
 The Chinese Academy of Sciences}\\
 \textit{P.O.Box 2735 Beijing 100080, China}} \date{}\maketitle

\section*{ABSTRACT}
The structures and symmetries of single-walled carbon nanotubes
(SWNTs) are introduced in detail. The physical properties of SWNTs
induced by their symmetries can be described by tensors in
mathematical point of view. It is found that there are 2, 4, and 5
different parameters in the second, third, and fourth rank tensors
representing electronic conductivity (or static polarizability),
the second order nonlinear polarizability, and elastic constants
of SWNTs, respectively. The values of elastic constants obtained
from tight-binding method imply that SWNTs might be very weakly
anisotropic in mechanical properties. The further study on the
mechanical properties shows that the elastic shell theory in the
macroscopic scale can be applied to carbon nanotubes (CNTs) in the
mesoscopic scale, as a result, SWNTs can be regarded as an
isotropic material with Poisson ratio, effective thickness, and
Young's modulus being $\nu=0.34$, $h=0.75$\AA, $Y=4.70$TPa,
respectively, while the Young's moduli of multi-walled carbon
nanotubes (MWNTs) are apparent functions of the number of layers,
$N$, varying from 4.70TPa to 1.04TPa for $N=1$ to $\infty$. Based
on the chirality of CNTs, it is predicted that a new kind of
molecular motor driven by alternating voltage can be constructed
from double walled carbon nanotubes (DWNTs).

\section*{INTRODUCTION}
Carbon is the core element to construction of organic matters and
has always attracted much attention up to now. Many decades ago,
people only knew two kinds of crystals consisting of carbon:
graphite with layer structure and diamond with tetrahedral shape.
The situation was changed in 1985 when Kroto \textit{et al.}
synthesized bucky ball---a football-like molecule consisting of 60
carbon atoms \cite{Kroto} which marked the beginning of carbon
times. After that, Iijima synthesized MWNTs in 1991 \cite{Iijima1}
and SWNTs in 1993 \cite{Iijima2}. Simply speaking, a SWNT can be
regarded as a graphitic sheet with hexagonal lattices that was
wrapped up into a seamless cylinder with diameter in nanometer
scale and length from tens of nanometers even to several
micrometers if we ignore its two end caps, while a MWNT consists
of a series of coaxial SWNTs with layer distance about 3.4\AA
\cite{Saito}.

SWNTs have many unique properties. Viewed from the chirality, some
of them are chiral but others are achiral. Viewed from the
electronic properties, some of them are metallic but others are
semiconductive. Moreover, their electronic properties depend
sensitively on their chirality \cite{Mintmire,Saito2,Hamada,Tans}.
The conductivity of metallic SWNT does not satisfy Ohm's law
because the electron transport in it is ballistic
\cite{Kong,Liang,White}. Otherwise, theoretical \cite{bi,x,e,jp}
and experimental \cite{a} studies have suggested that SWNTs also
possess many novel mechanical properties, in particular high
stiffness and axial strength, which are insensitive to the tube
diameters and chirality. MWNTs have the similar mechanical
properties to SWNTs \cite{mf,mmj,ew}.

In purely theoretical point of view, we should consider the
following two facts:

(i) As quasi-one-dimensional structures with periodic boundary
conditions, SWNTs might show anisotropic physical properties which
may depend on the tube diameters and chirality. Generally
speaking, the physical properties of crystals can be represented
by tensors \cite{Nyejf}. For example, the electronic conductivity
can be expressed by second-rank tensor and the elastic constants
can be described as fourth-rank tensor. These tensors can be
derived from the structures and symmetries of SWNTs. Therefore
fully discussing structures and symmetries of SWNTs is one of the
topic in this chapter.

(ii) The SWNT is a single layer of carbon atoms. What is the
thickness of the layer? It is a widely controversial question.
Some researchers take 3.4 \AA, the layer distance of bulk
graphite, as the thickness of SWNT \cite{jp,a}. Others define an
effective thickness (about 0.7\AA) by admitting the validity of
elastic shell theory in nanometer scale \cite{bi,x}. The present
authors have proved that the elastic shell theory can indeed be
applied to SWNTs \cite{tzcnt1} which supports the latter
standpoint. Because the two standpoints have little effect on the
mechanical results in many cases, the controversy is still being
discussed
\cite{Tabar,Pantano,Chandra,Ogata,Pantano2,Natsuki,SunCQ}.
Therefore it is necessary to point out when the two standpoints
will give different results.

In the applied point of view, the present authors have predicted a
molecular motor constructed from a DWNT driven by temperature
variation \cite{tzcnt2} which induces to molecular dynamics
simulations by Dendzik \textit{et al.} \cite{Dendzik}. Using
molecular dynamics simulations, Kang \textit{et al.} recently
predicted a carbon-nanotube motor driven by fluidic gas
\cite{KangJW2}. In this chapter, a conceptual motor of DWNT driven
by alternating voltage will be proposed based on previous work
\cite{tzcnt2}.

\section*{STRUCTURES, SYMMETRIES AND THEIR INDUCING PHYSICAL PROPERTIES OF SWNTS}
\begin{figure}[!htp]
\begin{center}\includegraphics[width=7cm]{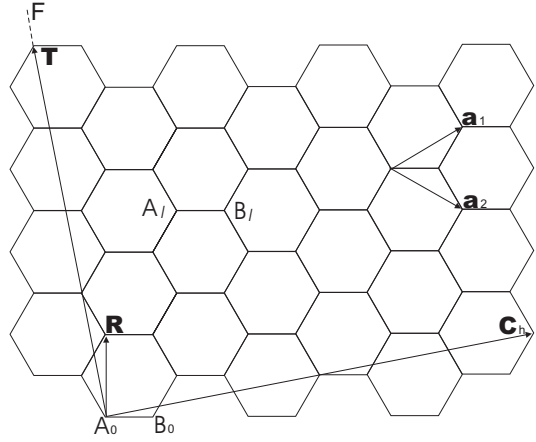}\end{center}
\caption{\label{fig1}The unrolled honeycomb lattice of a SWNT. By
rolling up the sheet along the chiral vector ${\bf C}_h$, that is,
such that the point $A_0$ coincides with the point corresponding
to vector ${\bf C}_h$, a nanotube is formed. The vectors ${\bf
a}_{1}$ and ${\bf a}_{2}$ are the real space unit vectors of the
hexagonal lattice. The translational vector ${\bf T}$ is
perpendicular to ${\bf C}_h$ and runs in the direction of the tube
axis. The vector ${\bf R}$ is the symmetry vector. $A_0$, $B_0$
and $A_l, B_l (l=1,2,\cdots,N)$ are used to denote the sites of
carbon atoms.}
\end{figure}

To describe the SWNT, some characteristic vectors require
introducing. As shown in Fig.\ref{fig1}, the chiral vector ${\bf
C}_{h}$, which defines the relative location of two sites, is
specified by a pair of integers $(n, m)$ which is called the index
of the SWNT and relates ${\bf C}_{h}$ to two unit vectors ${\bf
a}_{1}$ and ${\bf a}_{2}$ of graphite (${\bf C}_{h}=n{\bf
a}_{1}+m{\bf a}_{2}$). The chiral angle $\theta_0$ defines the
angle between $\mathbf{a}_1$ and $\mathbf{C}_h$. For $(n, m)$
nanotube,
$\theta_0=\arccos\left[\frac{2n+m}{2\sqrt{n^2+m^2+nm}}\right]$.
The translational vector ${\bf T}$ corresponds to the first
lattice point of 2D graphitic sheet through which the line normal
to the chiral vector ${\bf C}_{h}$ passes. The unit cell of the
SWNT is the rectangle defined by vectors ${\bf C}_{h}$ and ${\bf
T}$, while vectors ${\bf a}_{1}$ and ${\bf a}_{2}$ define the area
of the unit cell of 2D graphite. The number $N$ of hexagons per
unit cell of SWNT is obtained as a function of $n$ and $m$ as
$N=2(n^2+m^2+nm)/d_R$ which is larger than 8 for SWNTs in
practice, where $d_R$ is the greatest common divisor of ($2m+n$)
and ($2n+m$). There are $2N$ carbon atoms in each unit cell of
SWNT because every hexagon contains two atoms. To denote the $2N$
atoms, we use a symmetry vector ${\bf R}$ to generate coordinates
of carbon atoms in the nanotube and it is defined as the site
vector having the smallest component in the direction of ${\bf
C}_h$. From a geometric standpoint, vector ${\bf R}$ consists of a
rotation around the nanotube axis by an angle $\Psi=2\pi/N$
combined with a translation $\tau$ in the direction of ${\bf T}$;
therefore, ${\bf R}$ can be denoted by ${\bf R}=(\Psi|\tau)$.
Using the symmetry vector ${\bf R}$, we can divide the $2N$ carbon
atoms in the unit cell of SWNT into two classes \cite{tzcnt3}: one
includes $N$ atoms whose site vectors satisfy
\begin{equation}\label{sitea}
{\bf A}_l=l{\bf R}-[l{\bf R}\cdot{\bf T}/{\bf T}^2]{\bf T} \quad
(l=0,1,2,\cdots,N-1),\end{equation} another includes the remainder
$N$ atoms whose site vectors satisfy
\begin{eqnarray}\label{siteb}&&{\bf B}_l=l{\bf R}+{\bf
B}_0-[(l{\bf R}+{\bf B}_0)\cdot{\bf T}/{\bf T}^2]{\bf T}\nonumber
\\&&\quad -[(l{\bf R}+{\bf B}_0)\cdot{\bf C}_h/{\bf C}_h^2]{\bf C}_h
\quad (l=0,1,\cdots,N-1),\end{eqnarray}
 where ${\bf B}_0\equiv\left (\Psi_0|\tau_0\right)=\left (\left.\frac{2\pi
r_0\cos(\theta_0-\frac{\pi}{6})}{|\mathbf{C}_h|}\right|r_0\cos(\theta_0-\frac{\pi}{6})\right)$
represents one of the nearest neighbor atoms to {\bf A}$_0$ and
$r_0$ is the carbon-carbon bond length.

\begin{figure}[!htp]
\begin{center}\includegraphics[width=5cm]{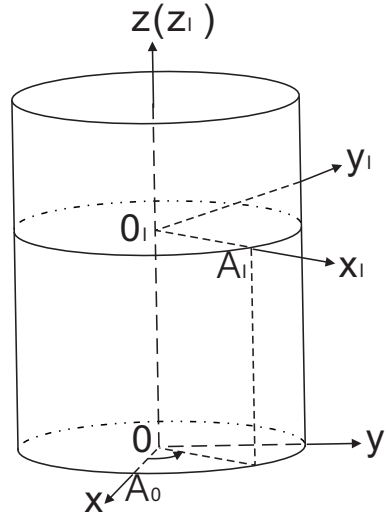}\end{center}
\caption{\label{fig2} The coordinates of SWNT.}
\end{figure}

If we introduce cylindrical coordinate system $(r, \theta, z)$
whose $z$-axis is the tube axis parallel to vector $\mathbf{T}$.
Its $r\theta$-plane is perpendicular to $z$-axis and contains atom
$A_0$ in the nanotube. $r$ the distance from some point to
$z$-axis, and $\theta$ the angle rotating around $z$-axis from an
axis which is vertical to $z$-axis and passes through atom $A_0$
in the tube to the point. In this coordinate system, we can
express Eqs.(\ref{sitea}) and (\ref{siteb}) as \cite{tzcnt33}:
\begin{equation}\label{siteaa}
{\bf A}_l=\{\rho,l\Psi,l\tau-[l\tau/T]T\}\quad
(l=0,1,2,\cdots,N-1) ,\end{equation} and
\begin{eqnarray}\label{sitebb} {\bf
B}_l&=&\left\{\rho,l\Psi+\Psi_0-2\pi\left[\frac{l\Psi+\Psi_0}{2\pi}\right],l\tau+\tau_0-\left[\frac{l\tau+\tau_0}{T}\right]T\right\}\nonumber\\
&&\quad (l=0,1,2,\cdots,N-1),\end{eqnarray} where
$\rho=\frac{|\mathbf{C}_h|}{2\pi}$. In
Eqs.(\ref{sitea})-(\ref{sitebb}), the symbol $[\cdots]$ denotes
the largest integer smaller than $\cdots$, e.g., $[7.3]=7$.

In the following contents of this section, we will derive the
general forms of the second, third, and fourth rank tensors by
fully considering the symmetries of SWNTs. As shown in
Fig.\ref{fig2}, $oxyz$ is the initial coordinate system whose
$z$-axis is the tube axis and $x$-axis passes through atom $A_0$.
If we use ${\bf R}$ to act $l$ times on the initial system, we can
get the coordinate system $O_l x_l y_l z_l$. Obviously, the
transformation from the bases $\hat{x}, \hat{y}, \hat{z}$ of the
initial system to the new bases $\hat{x}_l, \hat{y}_l, \hat{z}_l$
can be expressed as,
\begin{equation} \label{trans}
\left ( \begin{array}{c}
\hat{x}_{l}\\
\hat{y}_{l}\\
\hat{z}_{l}
\end{array} \right )=(a_{ij})
\left ( \begin{array}{c} \hat{x}\\ \hat{y}\\ \hat{z}\\
\end{array} \right )
,\end{equation} where $(a_{ij})$ is the matrix with elements
$a_{11}=a_{22}=\cos l\psi$, $a_{12}=\sin l\psi$, $a_{21}=-\sin
l\psi$, $a_{33}=1$, $a_{13}=a_{31}=a_{23}=a_{32}=0$.

Above all, let us consider the second-rank tensor ${\bf S}$.
$s_{ij}$ and $s_{ij}^{(l)}$ $(i,j=1,2,3; l=1,2,\cdots,N)$ denote
its components in the coordinate system $Oxyz$ and $O_l x_l y_l
z_l$. The transformation law of the components is
$s_{ij}^{(l)}=a_{im}a_{jk}s_{mk}$ \cite{Nyejf}, where the Einstein
summation convention is used. But the symmetry of SWNTs requires
$s_{ij}^{(l)}=s_{ij}$. From this condition and the transformation
law we derive out $s_{11}=s_{22}$, $s_{12}=-s_{21}$,
$s_{13}=s_{31}=0$, $s_{23}=s_{32}=0$. Especially, there are only 2
different nonzero parameters in symmetric 2nd-rank tensors which
represent the electronic conductivity or static polarizabilities:
$s_{11}=s_{22}\neq 0$, $s_{33}\neq 0$.

Next, let us deal with the third-rank tensor ${\bf D}$ whose
components are denoted by $d_{ijk}$ and $d_{ijk}^{(l)}$
$(i,j,k=1,2,3)$ in the coordinate system $Oxyz$ and $O_l x_l y_l
z_l$, respectively. From the transformation law of the components
$d_{ijk}^{(l)}=a_{iq}a_{ju}a_{kv}d_{quv}$ \cite{Nyejf} and the
symmetry of SWNTs, we can obtain the non-vanishing components of
${\bf D}$: $d_{113}=d_{223}$, $d_{123}=-d_{213}$,
$d_{131}=d_{232}$, $d_{132}=-d_{231}$, $d_{311}=d_{322}$,
$d_{312}=-d_{321}$, $d_{333}$. If some physical property requires
$d_{ijk}=d_{ikj}$ (e.g. the second order polarization effect),
then there are only 4 different nonzero parameters in its
components: $d_{123}=-d_{213}=d_{132}=-d_{231}\equiv d_1/2$,
$d_{113}=d_{223}=d_{131}=d_{232}\equiv d_2/2$,
$d_{311}=d_{322}\equiv d_3$, $d_{333}\equiv d_4$. Thus the
relation $P_i^{NL}=d_{ijk}E_jE_k$ between the nonlinear
polarization and field can be expressed in matrix form,
\begin{equation} \label{nonpolar}
\left\{ \begin{array}{l}
P_x^{NL}=d_1 E_yE_z+d_2 E_xE_z\\
P_y^{NL}=d_2 E_yE_z-d_1 E_xE_z\\
P_z^{NL}=d_3 (E_x^2+E_y^2)+d_4 E_z^2
\end{array} \right.
.\end{equation}

Similarly, we can obtain the nonzero components of the fourth-rank
tensor ${\bf C}$ from the transformation law
$c_{ijkm}^{(l)}=a_{iq}a_{ju}a_{kv}a_{mw}c_{quvw} (i,j,k,m=1,2,3)$
\cite{Nyejf} and the symmetry of SWNTs: $c_{1111}=c_{2222}$,
$c_{1112}=-c_{2221}$, $c_{1121}=-c_{2212}$, $c_{1122}=c_{2211}$,
$c_{1133}=c_{2233}$, $c_{1211}=-c_{2122}$, $c_{1212}=c_{2121}$,
$c_{1221}=c_{2112}=c_{2222}-c_{2121}-c_{2211}$,
$c_{2111}=-c_{1222}=c_{2122}+c_{2212}+c_{2221}$,
$c_{1233}=-c_{2133}$, $c_{1313}=c_{2323}$, $c_{1323}=-c_{2313}$,
$c_{1331}=c_{2332}$, $c_{1332}=-c_{2331}$, $c_{3113}=c_{3223}$,
$c_{3123}=-c_{3213}$, $c_{3131}=c_{3232}$, $c_{3132}=-c_{3321}$,
$c_{3311}=c_{3322}$, $c_{3312}=-c_{3321}$, $c_{3333}$. If we
consider the elastic property, the elastic constants can be
expressed by the fourth-rank tensor whose components satisfy
$c_{ijkm}=c_{ijmk}=c_{jikm}=c_{kmij}$\cite{Nyejf}. Thus there are
only 5 different non-vanishing parameters ($c_1, c_2, c_3, c_4,
c_5$) in its components and $c_{1111}=c_{2222}\equiv c_1$,
$c_{1133}=c_{3311}=c_{2233}=c_{3322}\equiv c_2$, $c_{3333}=c_3$,
$c_{1313}=c_{2323}=c_{1331}=c_{2332}=c_{3113}=c_{3223}=c_{3131}=c_{3232}\equiv
c_4$, $c_{1122}=c_{2211}\equiv c_5$,
$c_{1212}=c_{2121}=c_{1221}=c_{2112}=(c_1-c_5)/2$. The
stress-strain relation ${\bm \sigma}={\bf C}{\bm \varepsilon}$ can
be expressed by the matrix notations,
\begin{equation} \label{c2}
\left(\begin{array}{c} \sigma_{xx} \\ \sigma_{yy} \\ \sigma_{zz}
\\  \sigma_{yz} \\ \sigma_{xz} \\ \sigma_{xy}
\end{array} \right)=\left(\begin{array}{cccccc}
c_1 &  c_5 & c_2 & & &  \\
c_5 &  c_1 & c_2 & & &  \\
c_2 &  c_2 & c_3 & & &  \\
 & & & c_4& & \\
 & & & &c_4 & \\
 & & & & &\frac{(c_1-c_5)}{2}
\end{array} \right) \left( \begin{array}{c}
\varepsilon_{xx} \\ \varepsilon_{yy} \\ \varepsilon_{zz} \\
\gamma_{yz} \\ \gamma_{xz} \\ \gamma_{xy}
\end{array} \right)
,\end{equation} where $\gamma_{xy}=2\varepsilon_{xy}$,
$\gamma_{xz}=2\varepsilon_{xz}$,$\gamma_{yz}=2\varepsilon_{yz}$
\cite{ld}. The axial Young's modulus $Y_z$ defined as the
stress/strain ratio when the tube is axially strained and Poisson
ratio $\nu_z$ defined as the ratio of the reduction in radial
dimension to the axial elongation can be expressed as
$Y_z=c_3-2c_2^2/(c_1+c_5)$ and $\nu_z=c_2/(c_1+c_5)$,
respectively.

Obviously, the numbers of different parameters in the expressions
{\bf S}, {\bf D}, {\bf C} of SWNTs are more than that in isotropic
materials (see also Table \ref {number}) if the different
parameters are independent, which implies that SWNTs might possess
anisotropic physical properties. All parameters are functions of
$n$, $m$, which reveals the physical properties depend on the
chirality and diameters of SWNTs to some extent.
\begin {table}[htbp!]
\caption{\label{number}The non-zero parameter numbers of different
rank physical property tensors for isotropic materials and
single-walled carbon nanotubes.}
\begin{center}
\begin {tabular} {|cccc|}\hline
Tensors & {\bf S}(2nd-rank) &{\bf D}(3rd-rank)& {\bf C}(4th-rank) \\
\hline
Isotropic materials& 1 & 0 & 2 \\
carbon nanotubes & 2 & 4 & 5 \\
\hline
\end{tabular}
\end {center}
\end{table}

As examples, we will give the forms of second and fourth-rank
order tensors---the static polarizabilities and elastic constants
of SWNTs, respectively.

It is well known that the relation between the polarization {\bf
P} and external electric field {\bf E} is ${\bf P}={\bm
\alpha}\bf{E}$\cite{Landau2}, where $\bm{\alpha}$ is the static
polarizability, a second-rank tensor. From above discussions, we
known $\bm{\alpha}$ can be expressed as matrix form,
\begin{equation} \label{polar}
\bm{\alpha}=\left(\begin{array}{ccc}
\alpha_{xx} &  &  \\
 &\alpha_{yy} &  \\
 & & \alpha_{zz}
\end{array} \right)
,\end{equation} with $\alpha_{xx}=\alpha_{yy}$. Benedict {\it et
al.}\cite{Benedict} have studied the polarizabilities and their
results are shown in Table \ref{lxt}, where we have changed their
values to polarizabilities per atom. From Table \ref{lxt} we find
that the polarizabilities of SWNTs are sensitive to the tube
indexes $(n,m)$, particularly, the value of $\alpha_{zz}$ is
extremely large when $n-m$ is a multiple of three, which
corresponds metallic tubes.
\begin {table}[htbp!]
\caption{\label{lxt}Static polarizabilities per atom of various
tube indexes and radii \cite{Benedict}. When $n-m$ is multiple of
three, $\alpha_{zz}$ is extremely large and is not given.}
\begin{center}
\begin {tabular} {|cccc|}
\hline
$(n,m)$ & $\rho$(\AA) &$\alpha_{zz}$(\AA$^3$/atom) & $\alpha_{xx}$(\AA$^3$/atom)  \\
\hline
(9,0)& 3.57 &  & 1.05 \\
(10,0) & 3.94 & 18.62 & 1.10 \\
(11,0) & 4.33 & 17.04 & 1.20 \\
(12,0) & 4.73 & & 1.23\\
(13,0) & 5.12 & 23.98 & 1.29\\
\hline
(4,4) & 2.73 & & 0.92\\
(5,5) & 3.41 & & 1.02\\
(6,6) & 4.10 & & 1.13\\
\hline
(4,2) & 2.09 & 9.87 & 0.84\\
(5,2) & 2.46 & & 0.89\\
\hline
\end{tabular}
\end {center}
\end{table}

Otherwise, we calculate the elastic constants of SWNTs through the
tight binding method \cite{x} with considering the curvature and
bond-length change effects. The results are listed in Table
\ref{const} which suggests that the elastic properties of SWNTs
slightly depend on the tube indexes $(n,m)$. We also give the
corresponding axial Young's moduli and Poisson ratios of
single-walled carbon nanotubes with different indexes. Moreover,
we find $(c_1+c_5)-c_4\approx c_2$, $c_3\approx 2(c_1+c_5)$ and
$(c_1+c_5)\approx 4c_2$, i,e., there might be only two independent
parameters in the elastic constants, which implies that the
mechanical anisotropy of SWNTs is so weakly that we can regard
them as approximately isotropic materials (Remark: the isotropic
materials have two independent elastic constants, see also Table
\ref{number}).

\begin {table}[htbp!]
\caption{\label{const}The elastic constants (unit: eV/atom) and
corresponding axial Young's moduli (unit: eV/atom), Poisson ratios
of single-walled carbon nanotubes.}
\begin{center}
\begin {tabular} {|ccccccc|}
\hline
 (n,m)   &    $c_1+c_5$    &  $c_2$   &   $c_3$  &    $c_4$   &
$Y_z$ &
$\nu_z$\\
 \hline
(6,0) & 29.04 &  7.03 & 56.97 & 22.72 & 53.56 &  0.24 \\
(8,0) & 29.34 &  7.05 & 57.68 & 23.43 & 54.29 &  0.24 \\
(10,0) & 29.50 &  7.07 & 57.92 & 23.79 & 54.53 &  0.24 \\
(50,0) & 29.51 &  7.08 & 58.97 & 24.36 & 55.57 &  0.24 \\
\hline
(6,6) & 29.99 &  7.08 & 56.97 & 24.09 & 53.63 &  0.24 \\
(8,8) & 29.76 &  7.08 & 57.92 & 24.22 & 54.55 &  0.24 \\
(10,10) & 29.66 &  7.08 & 58.33 & 24.28 & 54.96 &  0.24 \\
(50,50) & 29.51 &  7.08 & 59.00 & 24.38 & 55.60 &  0.24 \\
\hline
(6,4) & 30.21 &  7.07 & 56.15 & 23.13 & 52.83 &  0.23 \\
(7,3) & 30.00 &  7.07 & 56.60 & 22.53 & 53.27 &  0.24 \\
(8,2) & 29.64 &  7.07 & 57.44 & 22.62 & 54.07 &  0.24 \\
 \hline
\end{tabular}\end{center}
\end{table}

It is necessary to discuss the meanings of strains and stresses in
nanometer scale. Strains are geometric quantities so that their
definitions in macroscopic theory of elasticity still hold for
SWNTs. But we must redefine stresses because they are not
well-defined quantities for SWNTs. Given strains, we can calculate
the energy variation of the SWNTs due to the strains through
quantum mechanics in principle. The stresses are defined as the
partial derivatives of energy variation with respect to the
strains. In fact, stresses are not necessary concepts. We can
directly determine the elastic constants by strains and the
corresponding energy variation.

Otherwise, we do not separate $c_1$ and $c_5$ in Table
\ref{const}. Up to now, we do not know how to separate them. A
possibility is that we need not do that when we discuss the
mechanical properties of SWNTs.

\section*{MECHANICAL PROPERTIES OF CNTS}
In this section, we will continue to discuss the mechanical
properties of CNTs in detail.

We start from the concise formula proposed by Lenosky {\it et al.}
in 1992 to describe the deformation energy of a single layer of
curved graphite\cite{Lenosky}
\begin{eqnarray} \label{leno}
E^g&=&(\epsilon_{0}/2) \sum_{(ij)}\left(r_{ij}-r_{0}\right)
^2+\epsilon_{1} \sum_{i}(\sum_{(j)}{\bf u}_{ij})^2 \nonumber \\
&+& \epsilon_{2} \sum_{(ij)}\left(1-{\bf n}_{i} \cdot {\bf n}_{j}
\right)+\epsilon_{3} \sum_{(ij)}\left({\bf n}_{i} \cdot {\bf
u}_{ij} \right) \left({\bf n}_{j} \cdot {\bf u}_{ji} \right)
.\end{eqnarray} The first two terms are the contributions of bond
length and bond angle changes to the energy. The last two terms
are the contributions of the $\pi$-electron resonance. In the
first term, $r_{0}=1.42$ \AA \ \ is the initial bond length of
planar graphite, and $r_{ij}$ is the bond length between atoms $i$
and $j$ after the deformations. In the remaining terms, ${\bf
u}_{ij}$ is a unit vector pointing from atom $i$ to its neighbor
$j$, and ${\bf n}_{i}$ is the unit vector normal to the plane
determined by the three neighbors of atom $i$. The summation
${\sum_{(j)}}$ is taken over the three nearest neighbor $j$ atoms
to $i$ atom, and ${\sum_{(ij)}}$ taken over all the nearest
neighbor atoms. The parameters
$(\epsilon_{1},\epsilon_{2},\epsilon_{3})=(0.96,1.29,0.05)$eV were
determined by Lenosky {\it et al.} \cite{Lenosky} through local
density approximation. The value of $\epsilon_{0}$ was not given
by Lenosky {\it et al.}, but given by Zhou {\it et al.}
\cite{zhoux} $\epsilon_{0}=57{\rm eV/\AA^2}$ from the
force-constant method.

In 1997, Ou-Yang \textit{et al.} \cite{oy1997} reduced
Eq.(\ref{leno}) into a continuum limit form without taking the
bond length change into account and obtained the curvature elastic
energy of a SWNT
\begin{equation} \label{oy1}
E^{(s)}=\int \left[\frac{1}{2} k_{c} (2H)^2+\bar{k}_{1} K \right]
dA ,\end{equation} where the bending elastic constant
\begin{equation} \label{oy2}
k_{c}=(18\epsilon_{1}+24\epsilon_{2}+9\epsilon_{3})r_{0}^2/(32
\Omega)=1.17{\rm eV}
\end{equation}
with $\Omega=2.62$ \AA$^2$\, being the occupied area per atom, and
\begin{equation} \label{oy3}
\bar{k}_{1}/k_{c}=-(8\epsilon_{2}+3\epsilon_{3})/(6\epsilon_{1}+8\epsilon_{2}+3\epsilon_{3})=-0.645
.\end{equation} In Eq.(\ref{oy1}), $H$, $K$, and $dA$ are mean
curvature, Gaussian curvature, and area element of the SWNTs
surface, respectively.

In 2002, we obtained the total free energy \cite{tzcnt1} of a
strained SWNT with in-plane strain ${\bf
\varepsilon}_i=\left(\begin{array}{cc} \varepsilon_{x} &
\varepsilon_{xy} \\ \varepsilon_{xy} & \varepsilon_{y} \end{array}
\right)$ at the $i$-atom site, where $\varepsilon_{x}$,
$\varepsilon_{y}$, and $\varepsilon_{xy}$ are the axial,
circumferential, and shear strains, respectively. The total free
energy contains two parts: one is the curvature energy expressed
as Eq.(\ref{oy1}); another is the deformation energy \cite{tzcnt1}
\begin{equation} \label{tzc20021}
E_{d}=\int \left[\frac{1}{2} k_{d} (2J)^2+\bar{k}_{2} Q \right] dA
,\end{equation} where $2J=\varepsilon_{x}+\varepsilon_{y}$ and
$Q=\varepsilon_{x} \varepsilon_{y}-\varepsilon_{xy}^2$, are
respectively named ``mean'' and ``Gaussian'' strains, and
\begin{equation} \label{tzc20022}
k_{d}=9\left(\epsilon_{0}
r_{0}^2+\epsilon_{1}\right)/(16\Omega)=24.88{\rm eV/\AA}^2
,\end{equation}
\begin{equation} \label{tzc20023}
\bar{k}_{2}=-3\left(\epsilon_{0}
r_{0}^2+3\epsilon_{1}\right)/(8\Omega)=-0.678k_d.\end{equation}
The value of $\bar{k}_{2}/k_d$ is so excellently close to the
value of $\bar{k}_{1}/k_{c}$ shown in Eq.(\ref{oy3}) that we can
regard that they are, in fact, equal to each other. We assume both
$\bar{k}_{2}/k_{d}$ and $\bar{k}_{1}/k_{c}$ are equal to their
average value,
\begin{equation} \label{tzc200266}
\bar{k}_{1}/k_{c}=\bar{k}_{2}/k_{d}=-0.66 .\end{equation} This is
the key relation that allows to describe the deformations of SWNT
with classic elastic theory. Thus, the deformation energy of a
SWNT, the sum of Eqs.(\ref{oy1}) and (\ref{tzc20021})
\begin{equation}
\label{tzc20024} E_{d}^{(s)}=\int \left[\frac{1}{2} k_{c}
(2H)^2+\bar{k}_{1} K \right] dA+\int \left[\frac{1}{2} k_{d}
(2J)^2+\bar{k}_{2} Q \right] dA \end{equation} can be expressed as
the form of the classic shell theory\cite{ld}:
\begin{eqnarray} \label{bi19971}
E_{c}&=&\frac{1}{2}\int D\left[ (2H)^2-2(1-\nu)K \right] dA
\nonumber \\ &+&\frac{1}{2}\int
\frac{C}{1-\nu^2}\left[(2J)^2-2(1-\nu)Q \right] dA ,\end{eqnarray}
where $D=(1/12)Yh^3/(1-\nu ^2)$ and $C=Yh$ are bending rigidity
and in-plane stiffness of shell. $\nu$ is the Poisson ratio and
$h$ is the thickness of shell. Comparing Eq.(\ref{tzc20024}) with
Eq.(\ref{bi19971}), we have
\begin{equation} \label{tzcc2002}
\left \{ \begin{array}{l}
(1/12)Yh^3/(1-\nu^2)=k_c\\
Yh/(1-\nu^2)=k_d\\
1-\nu=-\bar{k}_{1}/k_{c}=-\bar{k}_{2}/k_{d}
\end{array} \right..\end{equation}
From above equations we obtain the Poisson ratio, effective wall
thickness, and Young's modulus of SWNTs are $\nu=0.34$,
$h=0.75$\AA \ \ and, $Y=4.70$TPa, respectively. Our numerical
results are close to those given by Yakobson {\it et al.}
\cite{bi}.

Through above discussion, we can declare that: (i)
Eqs.(\ref{tzc200266}), (\ref{tzc20024}) and (\ref{bi19971}) imply
that elastic shell theory in macroscopic scale can be applied to
the SWNT in mesoscopic scale provided that its radius are not too
small. (ii) SWNT can be regard as being made from isotropic
materials with $Y=4.70$TPa and $\nu=0.34$. Its effective thickness
can be well-defined as $h=0.75$\AA.

We now turn to discuss the axial Young's modulus of MWNT. A MWNT
can be thought of as a series of coaxial SWNTs with layer distance
$d=3.4{\rm \AA}$.

Due to the deformation energy of SWNT, we can write the
deformation energy of MWNT as \cite{tzcnt1}
\begin{eqnarray}
E^{(m)}&=&\sum_{l=1}^N\int \left[\frac{1}{2} k_{d}
(2J)^2+\bar{k}_{2} Q \right] dA\nonumber\\&+&\sum_{l=1}^N \pi
k_{c} L/ \rho_{l} +\sum_{l=1}^{N-1}(\Delta E_{coh}/d) \pi L
(\rho_{l+1}^2-\rho_l^2) ,\label{oy19974}\end{eqnarray} where
$\rho_{l}$ is the radius of the $l$-th layer from inner one, $N$
is the layer number of MWNT, and $\Delta E_{coh}=-2.04 {\rm
eV/nm^2}$ \cite{Girifalco} being the interlayer cohesive energy
per area of planar graphite. $L$ is the length of MWNT. The second
term in Eq.(\ref{oy19974}) expresses the summation of curvature
energies on all layers given in Eq.(\ref{oy1}), and the third term
represents the total interlayer cohesive energy which actually
arises from the relatively weaker Van der Waals' interactions. On
this account, we can reasonably believe that the axial strain
$\varepsilon_{x}$ and circumferential strain $\varepsilon_{y}$
still satisfy $\varepsilon_{y}=-\nu \varepsilon_{x}$ for every
layer of SWNT in the MWNT when uniform stresses apply along axial
direction. Thus Eq.(\ref{oy19974}) becomes
\begin{equation}E^{(m)}=\frac{k_{d}}{2}(1-\nu^2)\varepsilon_{x}^2 \sum_{l=1}^N
2\pi \rho_l L+\sum_{l=1}^N \pi k_{c} L/ \rho_{l}
+\sum_{l=1}^{N-1}\Delta E_{coh} \pi L (\rho_{l+1}+\rho_l)
.\label{oy19975}\end{equation}

The axial Young's modulus of the MWNT $Y_m$ is defined as
\begin{equation} \label{tzc20026}
Y_m(N)=\frac{1}{V}\frac{\partial^2 E^{(m)}}{\partial
\varepsilon_{x}^2} ,\end{equation} where $V=\pi L
[(\rho_N+h/2)^2-(\rho_1-h/2)^2]$ is the volume of MWNT. If
considering $\sum_{l=1}^N 2\pi \rho_l L=(\rho_1+\rho_N)N\pi L$,
from Eqs.(\ref{oy19975}) and (\ref{tzc20026}) we have
\cite{tzcnt1}
\begin{equation} \label{tzc20027}
Y_m(N)=\frac{N}{N-1+h/d}\frac{h}{d}Y ,\end{equation} where $Y$ and
$h$ are the Young's modulus and effective wall thickness of SWNTs.
Obviously, $Y_m=Y=4.70$TPa if $N$=1, which corresponds to the
result of SWNTs, and $Y_m=Yh/d=1.04$TPa if $N \gg 1$, which is
just the Young's modulus of bulk graphite. The layer number
dependence of MWNT's Young's modulus can be used to discuss
mechanical properties of nanotube/polymer composites
\cite{LauKT,LauShi,XiaoT}.

Now let us discuss the mechanical stabilities of SWNTs and MWNTs.
In above discussion, we have prove that elastic shell theory can
be applied to SWNTs. Thus we can directly use the classical
results in elastic shell theory \cite{wujk,Pogorelov}.

The critical pressure for SWNT by axial stress on both free ends
is
\begin{equation}\label{critpswnta}
p_{csa}=\frac{1}{\sqrt{3(1-\nu^2)}}\frac{Yh}{\rho}. \end{equation}
The critical pressure for SWNT (with two free ends) by radial
stress is \begin{equation}\label{critpswntr}
p_{csr}=\frac{Yh^3}{4(1-\nu^2)\rho^3}.\end{equation} The critical
moment for SWMT by uniform bending moment is
\begin{equation}
M_{cs}=\frac{2\sqrt{2}\pi Y\rho
h^2}{9\sqrt{1-\nu^2}}.\end{equation} The critical torsion for SWMT
by uniform torsion is
\begin{equation}
T_{cs}=\frac{\sqrt{2}\pi Y\sqrt{\rho
h^5}}{3(1-\nu^2)^{3/4}}.\end{equation} In above four equations,
$\nu=0.34$, $h=0.75$\AA, $Y=4.70$TPa, and $\rho$ is the radius of
SWNT.

The stability of MWNT is determined by its weakest layer of SWNT.
Thus we have the critical pressure for MWNT by axial stress on
both free ends \begin{equation}\label{critpmwnta}
p_{cma}=\frac{1}{\sqrt{3(1-\nu^2)}}\frac{Yh}{\rho_o}\end{equation}
and the critical pressure for MWNT (with two free ends) by radial
stress \begin{equation}\label{critpmwntr}
p_{cmr}=\frac{Yh^3}{4(1-\nu^2)\rho_o^3},\end{equation} where
$\nu=0.34$, $h=0.75$\AA, $Y=4.70$TPa, and $\rho_o$ is the outmost
radius of MWNT.

Eqs.(\ref{critpswnta})--(\ref{critpmwntr}) are valid provided that
the elastic shell theory can be applied to SWNTs. Adopting the
\textit{ad hoc} convention $h=3.4$\AA---the layer distance of bulk
graphite as the thickness of SWNTs will give different values of
these equations. Thus the crucial experiments to test above
relations will judge whether we should take the \textit{ad hoc}
convention or effective thickness of SWNTs.

\section*{MOTORS OF DWNTS}
\begin{figure}[htp!]\begin{center}
\includegraphics[width=7cm]{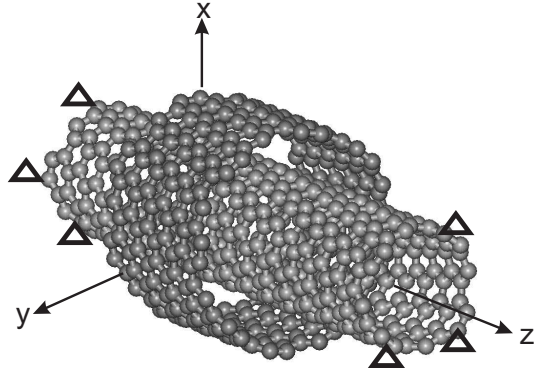}\end{center}
\caption{ A double-walled carbon nanotube with inner tube's index
being (8, 4) and outer tube's index being (14, 8). $z$-axis is the
tube axis parallel to vector $\mathbf{T}$. $x$-axis is
perpendicular to $z$ passes through one of carbon atoms in inner
tube and $y$-axis perpendicular to the $xz$-plane. The triangles
represent the fixed devices of outer tube. There is no obviously
relative motion along radial direction between inner and outer
tubes at low temperature. If we forbid the motion of outer tube in
the direction of $z$-axis, only the rotation of outer tube around
inner tube is permitted.}\label{dwnteps}
\end{figure}

With the development of nanotechnology, especially the discovery
of carbon nanotubes, people are putting their dream of
manufacturing nanodevices \cite{Feynman1,Drexler} into practice.
In this section, we will discuss the potential application of
CNTs. First, we specifically introduce a molecular motor
constructed from a DWNT driven by temperature variation due to the
different chirality of DWNT proposed in Ref.\cite{tzcnt2}. As
shown in Fig.\ref{dwnteps}, inner tube's index is (8, 4) with a
length long enough to be regarded as infinite while outer tube is
a (14, 8) tube with just a single unit cell \cite{footno1}.
Obviously, they are both chiral nanotubes and their layer distance
is about 3.4 \AA. If we prohibit the motion of outer tube in the
direction of nanotube axis, it will be proved that, in a thermal
bath, outer tube exhibits a directional rotation when the
temperature of the bath varies with time. Thus it could serve as a
thermal ratchet.

To see this, we first select an orthogonal coordinate system shown
in Fig.\ref{dwnteps} whose $z$-axis is the tube axis and $x$-axis
passes through one of carbon atoms in the inner tube. We fix the
inner tube and forbid the $z$-directional motion of the outer
tube. The rotation angle of outer tube around the inner one is
denoted by $\theta$.

\begin{figure}[htp!]
\begin{center}\includegraphics[width=6.8cm]{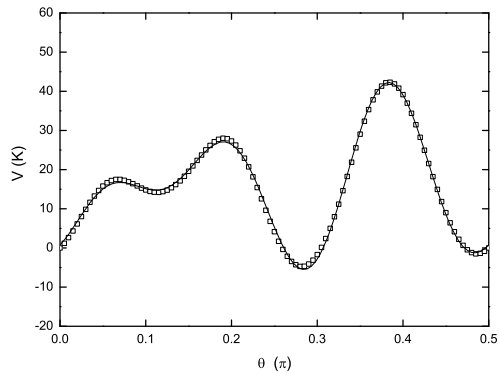}\end{center}
\caption{ The potentials $V(\theta)$ between outer and inner tubes
when outer tube rotates around inner tube. $\theta$ is the
rotating angle. Here we have set $V(0)=0$. The squares are the
numerical results which can be well fitted by
$V(\theta)=15.7-0.6\cos
4\theta-2.2\sin4\theta-12.7\cos8\theta-6\sin8\theta-1.7\cos12\theta+10.8\sin
12\theta$ (solid curve).}\label{potentialeps}
\end{figure}

We take the interaction between atoms in outer and inner tube as
the Lennard-Jones potential
$u(r_{ij})=4\epsilon[(\sigma/r_{ij})^{12}-(\sigma/r_{ij})^6]$,
where $r_{ij}$ is the distance between atom $i$ in inner tube and
atom $j$ in outer tube, $\epsilon=28$ K, and $\sigma=3.4$\AA\
\cite{hir}. We calculate the potential $V(\theta)$ when outer tube
rotates around inner tube with angle $\theta$ and plot it in
Fig.\ref{potentialeps}.

If putting our system in a thermal bath full of He gas whose
temperature varies with time, We can show that outer tube will
exhibit a directional rotation. Let us consider the overdamped
case when the Langevin equation for outer tube is expressed
\begin{equation}\label{Langevin}\eta\dot{\theta}=-V'(\theta) +\xi(t),\end{equation}
where $\eta$ is the rotating viscosity coefficient, and dot and
prime indicate, respectively, differentiations with respect to
time $t$ and angle $\theta$. $\xi(t)$ is thermal noise which
satisfies $\langle \xi(t)\rangle=0$ and the
fluctuation-dissipation relation $\langle\xi(t)\xi(s)\rangle=2\eta
T(t)\delta(t-s)$, where $T(t)$ is temperature and the Boltzmann
factor is set to 1. The Fokker-Planck equation corresponding to
Eq.(\ref{Langevin}) is \cite{Reimann}:
\begin{eqnarray}\label{fplanck}
\frac{\partial P(\theta,t)}{\partial
t}=\frac{\partial}{\partial\theta}\left[\frac{V'(\theta)P(\theta,t)}{\eta}\right]+\frac{T(t)}{\eta}\frac{\partial^2P(\theta,t)}{\partial\theta^2},
\end{eqnarray}
where $P(\theta,t)$ represents the probability of finding the
outer tube at angle $\theta$ and time $t$ which satisfies
$P(\theta+\pi/2,t)=P(\theta,t)$. If the period of temperature
variation is $\mathcal{T}$, we arrive at the average angular
velocity in the long-time limit \cite{Reimann}
\begin{equation}\label{current}
\langle\dot{\theta}\rangle=\lim_{t\rightarrow
\infty}\frac{1}{\mathcal{T}}\int_t^{t+\mathcal{T}}dt\int_0^{\pi/2}d\theta\left[-\frac{V'(\theta)P(\theta,t)}{\eta}\right].
\end{equation}

Here we take the periodical varying temperature
$T(t)=\bar{T}[1+A\sin(2\pi t/\mathcal{T})]$ with $\bar{T}=50$ K
and $|A|\ll 1$. Changing variables $D=\eta/\bar{T}$, $t=D\tau$,
$U(\theta)=V(\theta)/\bar{T}$, $\mathcal{T}=D\mathcal{J}$ and
$\tilde{P}(\theta,\tau)=P(\theta,D\tau)$, We arrive at the
dimensionless equations of Eqs.(\ref{fplanck}) and (\ref{current})
\begin{eqnarray}
&&\frac{\partial \tilde{P}}{\partial
\tau}=\frac{\partial}{\partial\theta}[U'(\theta)\tilde{P}]+(1+A\sin\frac{2\pi\tau}{\mathcal{J}})\frac{\partial^2\tilde{P}}{\partial\theta^2},\label{fplanck2}\\
&&\langle \frac{d\theta}{d\tau}\rangle=\lim_{\tau\rightarrow
\infty}\frac{1}{\mathcal{J}}\int_{\tau}^{\tau+\mathcal{J}}d\tau\int_0^{\pi/2}d\theta\left[-U'(\theta)\tilde{P}\right].\label{current2}
\end{eqnarray}

\begin{figure}[htp!]
\begin{center}\includegraphics[width=7cm]{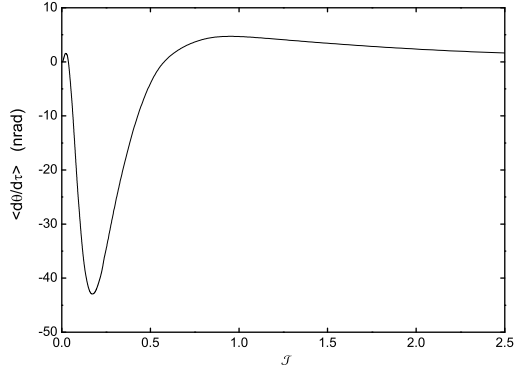}\end{center}
\caption{The average dimensionless angular velocity $\langle
d\theta/d\tau\rangle$ of outer tube rotating around inner tube in
thermal bath whose temperature changes with dimensionless period
$\mathcal{J}$. The minus sign means rotation around $z$-axis is
the left-handed, and vice versa. }\label{averagevel}
\end{figure}

We numerically solve Eq.(\ref{fplanck2}) and calculate
Eq.(\ref{current2}) with $A=0.01$ and different $\mathcal{J}$. The
curve in Fig.\ref{averagevel} shows the relation between the
average dimensionless angular velocity
$\langle\frac{d\theta}{d\tau}\rangle$ and the dimensionless period
$\mathcal{J}$ of temperature variation. We find that
$\langle\frac{d\theta}{d\tau}\rangle\simeq 0$ for very small and
large $\mathcal{J}$, and $\langle\frac{d\theta}{d\tau}\rangle\neq
0$ for the middle values of $\mathcal{J}$, which implies that
outer tube has an evident directional rotation in this period
range.

In above discussion, we have conceptually constructed a
temperature ratchet---a kind of molecular motor driven by
temperature variations which satisfies two necessary conditions
required by the second law of thermodynamics \cite{Feynman2}: One
is breaking spatial inversion symmetry (through different
chirality of two SWNTs in DWNT); Another is breaking of thermal
equilibrium (through temperature variations). But, in fact, this
motor is not convenient because it is hard to control temperature
variations. Our dream is construct an electric motor driven by
alternating voltage discussed as below.

Although no one believe SWNTs are piezoelectric materials, an
exceptionally large axial deformation in SWNTs induced by applied
electrostatic field along tube axis is demonstrated using
Hartree-Fock and density functional calculations by Guo \textit{et
al.} \cite{Guowl}. Additionally, their calculations reveal that
the bond-elongation of SWNTs is linearly dependent on the axial
field when the field is not strong enough. These results shed a
light on our dream of making the electric motor. Now, we will
qualitatively construct it. Put the device shown in
Fig.\ref{dwnteps} between two electrodes and apply alternating
voltage $U_A(t)=U_0\sin(\omega t)$ on the electrodes, where $U_0$
and $\omega$ are constants. We expect that the field due to
alternating voltage also induces bond-elongation of the DWNT. Thus
the layer distance will change, which will change the interaction
between layers. At least for small $U_0$, the lowest order Taylor
Series of interaction between layers might be written as
$\tilde{V}(\theta,t)\approx V(\theta)+\delta(\theta) |\sin\omega
t|$ with $\delta(\theta)$ being a small quantity. If putting our
system in a thermal bath full of He gas with constant temperature
$T$, we have the Langevin equation of outer tube in the overdamped
case \cite{Reimann}:
\begin{equation}\label{elctromotor}\eta\dot{\theta}=-[V'(\theta)+\delta'(\theta)
|\sin\omega t|] +\xi(t),\end{equation} where $\xi(t)$ is thermal
noise which satisfies $\langle \xi(t)\rangle=0$ and the
fluctuation-dissipation relation $\langle\xi(t)\xi(s)\rangle=2\eta
T\delta(t-s)$. Consequently, the thermal equilibrium is broken by
the shaking potential $\tilde{V}(\theta,t)$, and we may construct
a fluctuating potential ratchet proposed in
Ref.\cite{Reimann,RDAstumian}.

\section*{CONCLUSION}
In this chapter, we introduce the structures, symmetries and their
inducing physical properties of SWNTs in detail. We discuss the
mechanics of nanotubes and explain why we must define the
effective thickness of SWNTs. We also qualitatively construct an
electric motor driven by alternating voltage. We expect some
experiments which can put our electric motor into practice in near
future.

%%%%%%%%%%%%%%%%%%%%%%%%%%%%%%%%%%%%%%%%%%%%%%%%%%%%%%%%%%%%%%%%%%%%%%%%%%


\begin{thebibliography}{99}
\bibitem{footno1}If we consider several unit cells, the final qualitative results sitll hold.
\end{thebibliography}
\end{document}